\documentclass[12pt]{article}
\usepackage{amssymb}
\usepackage{graphicx}
\begin{document}
\null
\vskip 34mm

\centerline{{\bf A proposal for simulating QCD}}
\centerline{{\bf at finite chemical potential on the lattice}}
\vskip 5mm
\centerline{B.~All\'es, E.~M.~Moroni}
\centerline{\it Dipartimento di Fisica, Universit\`a di Milano-Bicocca}
\centerline{\it and INFN Sezione di Milano, Milano, Italy}
\begin{abstract}
An algorithm to simulate full QCD with 3 colours at nonzero chemical potential
on the lattice is proposed. The algorithm works for 
small values of the chemical potential and can be used to extract expectation
values of {\sf CPT} invariant operators.
\end{abstract}

\vfill\eject

\section{Introduction}

Understanding the properties of matter at nonzero density and temperature is an essential
ingredient to describe the physics of the early universe and the collapse of very massive
stars~\cite{rajagopalwnardulli}. Moreover recent heavy ion collision
experiments offer a unique laboratory test for the predictions of the theory~\cite{satzgavai}.

Effective models~\cite{rajagopalwnardulli,barroisbailincasalbuoni}
predict a rich structure in the temperature--chemical
potential $T$--$\mu$ diagram of QCD with several phases. Numerical
simulations on the lattice is an adequate technique to study the corresponding 
phase transitions.
The lattice action of QCD at finite $\mu$ has been given in~\cite{hasenfratzk,kogut}. By using 
Wilson fermions within a $SU(N)$ invariant gauge theory, this action is~\cite{wilson}
\begin{eqnarray}
S_{\rm Wilson}&\equiv& \beta \sum_{P} 
                       \left( 1 - {1\over N} \hbox{Re Tr}\, {P} \right) \nonumber \\
              & & +\sum_{\rm flavours} \sum_x a^3 \Big[ \left( am + 4 \lambda\right)
                   \overline\psi_x \psi_x \nonumber \\
              & &  \qquad\qquad\qquad - {1\over 2} \sum_\nu\Big(
                   \left(\lambda - \gamma_\nu\right) 
                   \overline\psi_{x} U_\nu(x) \xi_\nu(\mu) \psi_{x+\hat\nu} 
                   \nonumber \\
              & & \qquad\qquad\qquad + \left(\lambda + \gamma_\nu\right) 
                    \overline\psi_{x+\hat\nu} U^\dagger_\nu(x) \xi_\nu(\mu)^{-1} \psi_x
                    \Big)\Big]\; , \nonumber \\
              & & \nonumber \\
\xi_\nu(\mu) &\equiv& 1 + \delta_{4\nu} \left(\exp\left(f(a\mu)\right) - 1\right)\; ,
\label{action}
\end{eqnarray}
where the first term is the pure gauge action, $a$ is the lattice spacing,
$\beta$ is the inverse bare lattice coupling ($\beta=2N/g_0^2$), ${P}$ 
stands for the elementary plaquette, $m$ is the fermion mass, $\lambda$
is the Wilson parameter and $\mu$ is the chemical potential.
$f(a\mu)$ is an odd function of the chemical potential that satisfies
$f(x)= x + O(x^3)$. The simplest choice is $f(x)=x$ although others
are possible (see for instance~\cite{bilic} where $f(x)=\hbox{arg tanh} x$). 
Notice that $\xi_\nu(-\mu)=\xi_\nu(\mu)^{-1}$.
Analogous
expressions are valid for staggered and na\"{\i}ve fermions.

The expectation value of an operator $O$ is defined by the Feynman path integral
\begin{eqnarray}
\langle O \rangle &=& {1\over Z} \int {\cal D}U_\nu {\cal D}\overline\psi_x 
{\cal D}\psi_x \; O\,\exp\left(-S_{\rm Wilson}\right)\; ,  \nonumber \\
Z &\equiv& \int {\cal D}U_\nu {\cal D}\overline\psi_x 
{\cal D}\psi_x \; \exp\left(-S_{\rm Wilson}\right)\; ,
\label{vev}
\end{eqnarray}
where terms regarding gauge fixing have been skipped as they are inessential
for our analysis. After integrating out fermions  (we assume that $O$ can
be expressed in terms of gluon fields only) we have
\begin{equation}
\langle O \rangle = {1\over Z} \int {\cal D}U_\nu\; O\, \det D \; 
\exp\left(-S_{\rm g}\right) \; ,
\label{vevO}
\end{equation}
where $S_{\rm g}$ is the pure gauge action and $D$ is the fermion matrix
\begin{eqnarray}
D_{xy}(\mu)&=&(ma + \lambda) \delta_{x,y}\nonumber \\
           - {1\over 2} 
\negthinspace\negthinspace\negthinspace\negthinspace
&& \negthinspace\negthinspace\negthinspace\negthinspace
      \sum_\nu \left(\delta_{x,y+\hat\nu} U^\dagger_\nu(y) \left(\lambda +
      \gamma_\nu\right) \xi_\nu(\mu)^{-1} + 
      \delta_{y,x+\hat\nu} U_\nu(x) \left(\lambda -
      \gamma_\nu\right) \xi_\nu(\mu) \right)\; .
\label{D}
\end{eqnarray}
Colour, flavour and spinor indices are placed where required. The theory
with $N=3$ colours and fermions in the fundamental representation in general yields
a complex value for
$\det D$ if $\mu\not= 0$. This is a problem because importance sampling in the
Monte Carlo integration of~Eq.(\ref{vevO}) cannot be applied with a complex weight.
Several solutions have been proposed that allow to perform the integration in an
approximate way or in particular regions of the $T$--$\mu$ diagram: reweighting
methods~\cite{fodork}, calculation of the canonical partition function from 
simulations at imaginary chemical potential~\cite{alford}, analytical continuations of
Taylor expansions in powers of imaginary chemical potential~\cite{delial},
responses of several observables to a nonzero small chemical potential~\cite{hands}
and fugacity expansions~\cite{barbour}. Here we present
another idea that should work for small values of the chemical
potential. We will prove that if we restrict our attention to the calculation
of expectation values of {\sf CPT} invariant operators then the correct
Boltzmann weight under the path integral is 
$\exp\left(-S_{\rm g}\right)\hbox{Re} \det D$.

\section{The Boltzmann weight}

The Boltzmann weight in Eq.(\ref{vevO}),
$\exp\left(-S_{\rm g}\right) \det D$, is a complex number. However
the expectation value of any observable represented by the operator $O$ 
in~Eq.(\ref{vevO}) must be a real quantity. Imposing that $\langle O\rangle$ 
be real for any observable $O$ is a strong constraint on the integration measure.
Clearly all possible configurations can be gathered in sets such that the contribution
to the imaginary part of
$\langle O\rangle$ from all configurations in one single set cancels out.
We shall assume that all sets are formed by only two configurations
and that these two are related by some transformation {\sf S}. Firstly we want
to find this transformation.
Let ${\cal C}$ denote a thermalized arbitrary
configuration on the lattice and ${\cal C^{\sf S}}$ the corresponding transformed
configuration by the action of {\sf S}. We require that {\it (i)} the value of the
operator $O$ calculated on ${\cal C}$ and ${\cal C^{\sf S}}$ be the same; {\it (ii)}
the weight under the path integral for ${\cal C^{\sf S}}$ be the complex conjugate
of the weight for ${\cal C}$ and {\it (iii)} ${\cal C}$
and ${\cal C^{\sf S}}$ have the same Haar measure,
\begin{eqnarray}
O[{\cal C}] & \rightarrow & O[{\cal C^{\sf S}}] = O[{\cal C}]\;, \nonumber \\
\det D[{\cal C}] \hbox{e}^{-S_{\rm g}[{\cal C}]} 
& \rightarrow & \det D[{\cal C^{\sf S}}] \hbox{e}^{-S_{\rm g}[{\cal C^{\sf S}}]}
= \left(\det D[{\cal C}]\right)^* \hbox{e}^{-S_{\rm g}[{\cal C}]} \; , 
\nonumber \\
{\cal D}U_\nu[{\cal C}] & \rightarrow & {\cal D}U_\nu[{\cal C^{\sf S}}] = 
{\cal D}U_\nu[{\cal C}] \; ,
\label{conditions}
\end{eqnarray}
where we have explicitely written the dependence on the configuration.
We will write this dependence wherever necessary. These conditions are
sufficient to guarantee that $\langle O\rangle$ is real.

A simple way to enforce the last constraint in~(\ref{conditions})
is by imposing that {\sf S} is a discrete transformation. ${\cal D}U_\nu$ means
$\Pi_x \Pi_\nu \hbox{d}U_\nu(x)$ where each single factor $\hbox{d}U_\nu(x)$
is the Haar measure over the gauge group. By a discrete transformation
we mean that {\sf S} does not transform each single Haar measure. In fact
this would entail a jacobian and the search of the transformed configuration
would become more difficult.

The matrix $D_{xy}(\mu)$ has the property 
$\left(\det D(\mu)\right)^* = \det D(-\mu)$. This stems from the relation
$D(\mu)^\dagger=\gamma_5 D(-\mu)\gamma_5$\footnote{This is true
for Wilson and na\"{\i}ve fermions; for staggered fermions a matrix other than $\gamma_5$ 
must be used but the result is the same. We use the euclidean definition of
the gamma matrices such that $\gamma_\nu=\gamma_\nu^\dagger$ holds.}.
Notice that this implies
Re det$D$ (Im det$D$) is an even (odd) function of $\mu$.
Moreover it is physically clear that applying a charge conjugation operator
{\sf C} changes the sign of $\mu$. Then we expect that the condition 
$\det D[{\cal C^{\sf S}}] = \left(\det D[{\cal C}]\right)^*$ can be
verified if {\sf S} contains the transformation {\sf C}. However the
above condition is only part of the requirements in~(\ref{conditions}).
In particular we see from the second condition in~(\ref{conditions})
that the implementation of {\sf C} must be such that the pure gluon
action $S_{\rm g}$ remains unaltered.

In order to leave $S_{\rm g}$ invariant, the correct 
transformation should involve some space--time rearrangement besides 
charge conjugation. A reasonable guess is the 
{\sf CPT} transformation that we define in the following way:
if our lattice was $d$ dimensional and the lateral sizes were finite and equal
to $L_1$, $L_2$, ..., $L_d$ (in units of lattice spacing) then the {\sf CPT} 
transformation would act in the following way
\begin{eqnarray}
 U_\nu(x)^{{\sf CPT}} = U_\nu^\dagger( 
                           &&\negthinspace\negthinspace\negthinspace\negthinspace\negthinspace
                             \negthinspace\negthinspace\negthinspace\negthinspace
                             \hbox{mod}(L_1 - x_1 + 1,L_1) + 1,   \nonumber \\
                           && \negthinspace\negthinspace\negthinspace\negthinspace\negthinspace
                             \negthinspace\negthinspace\negthinspace\negthinspace
                             \hbox{mod}(L_2 - x_2 + 1,L_2) + 1,   \nonumber \\
                           && \negthinspace\negthinspace\negthinspace\negthinspace\negthinspace
                             \negthinspace\negthinspace\negthinspace\negthinspace
                             \cdots,                              \nonumber \\
                           && \negthinspace\negthinspace\negthinspace\negthinspace\negthinspace
                             \negthinspace\negthinspace\negthinspace\negthinspace
                              \hbox{mod}(L_\nu - x_\nu,L_\nu) + 1, \nonumber \\
                           && \negthinspace\negthinspace\negthinspace\negthinspace\negthinspace
                             \negthinspace\negthinspace\negthinspace\negthinspace
                             \cdots,                              \nonumber \\
                           && \negthinspace\negthinspace\negthinspace\negthinspace\negthinspace
                             \negthinspace\negthinspace\negthinspace\negthinspace
                             \hbox{mod}(L_d - x_d + 1,L_d) + 1 ) \; ,
\label{Ld}
\end{eqnarray}
where mod($x,L$) gives the remainder on integer division of $x$ by $L$.
This is the finite volume version of $U_\nu(x)^{\sf CPT}= U^\dagger_\nu(-x)$.
This transformation fulfils all conditions in~Eq.(\ref{conditions}) as a straightforward 
computation shows.

As an example let us check that $S_{\rm g}$ is invariant under {\sf CPT}.
It is enough to show that under the action of {\sf CPT} 
every plaquette of the original configuration ${\cal C}$
maps onto one and only one plaquette of ${\cal C}$. Let us denote ${P} (x;\nu,\rho)$
the plaquette starting at the site $x$ and going around through direction $\nu$ and
then $\rho$. This is ${P} (x;\nu,\rho)\equiv U_\nu(x) U_\rho(x+\hat\nu) U^\dagger_\nu(x+\hat\rho)
U^\dagger_\rho(x)$. Applying transformation~(\ref{Ld}) on each factor we obtain
\begin{eqnarray}
{P} (x;\nu,\rho) \rightarrow U^\dagger_\nu&&
\negthinspace\negthinspace\negthinspace\negthinspace\negthinspace
                             \negthinspace\negthinspace\negthinspace\negthinspace
(\hbox{mod}(L_1 - x_1+1, L_1)+1, \cdots \nonumber \\
                                           &&\negthinspace\negthinspace\negthinspace\negthinspace\negthinspace
                             \negthinspace\negthinspace\negthinspace\negthinspace
\hbox{mod}(L_\nu - x_\nu, L_\nu)+1, \cdots \nonumber \\
                            &&\negthinspace\negthinspace\negthinspace\negthinspace\negthinspace
                             \negthinspace\negthinspace\negthinspace\negthinspace
\hbox{mod}(L_\rho - x_\rho+1, L_\rho)+1, \cdots ) \times \nonumber \\
                              U^\dagger_\rho&&\negthinspace\negthinspace\negthinspace\negthinspace\negthinspace
                             \negthinspace\negthinspace\negthinspace\negthinspace
(\hbox{mod}(L_1 - x_1+1, L_1)+1, \cdots \nonumber \\
                                           &&\negthinspace\negthinspace\negthinspace\negthinspace\negthinspace
                             \negthinspace\negthinspace\negthinspace\negthinspace
\hbox{mod}(L_\nu - x_\nu, L_\nu)+1, \cdots \nonumber \\
                            &&\negthinspace\negthinspace\negthinspace\negthinspace\negthinspace
                             \negthinspace\negthinspace\negthinspace\negthinspace
\hbox{mod}(L_\rho - x_\rho, L_\rho)+1, \cdots ) \times \nonumber \\
                              U_\nu&&\negthinspace\negthinspace\negthinspace\negthinspace\negthinspace
                             \negthinspace\negthinspace\negthinspace\negthinspace
(\hbox{mod}(L_1 - x_1+1, L_1)+1, \cdots \nonumber \\
                                           &&\negthinspace\negthinspace\negthinspace\negthinspace\negthinspace
                             \negthinspace\negthinspace\negthinspace\negthinspace
\hbox{mod}(L_\nu - x_\nu, L_\nu)+1, \cdots \nonumber \\
                            &&\negthinspace\negthinspace\negthinspace\negthinspace\negthinspace
                             \negthinspace\negthinspace\negthinspace\negthinspace
\hbox{mod}(L_\rho - x_\rho, L_\rho)+1, \cdots ) \times \nonumber \\
                              U_\rho&&\negthinspace\negthinspace\negthinspace\negthinspace\negthinspace
                             \negthinspace\negthinspace\negthinspace\negthinspace
(\hbox{mod}(L_1 - x_1+1, L_1)+1, \cdots \nonumber \\
                                           &&\negthinspace\negthinspace\negthinspace\negthinspace\negthinspace
                             \negthinspace\negthinspace\negthinspace\negthinspace
\hbox{mod}(L_\nu - x_\nu +1, L_\nu)+1, \cdots \nonumber \\
                            &&\negthinspace\negthinspace\negthinspace\negthinspace\negthinspace
                             \negthinspace\negthinspace\negthinspace\negthinspace
\hbox{mod}(L_\rho - x_\rho, L_\rho)+1, \cdots )  \; , \nonumber\\
\end{eqnarray}
which, by the cyclic property of the trace, is equivalent to the plaquette
\begin{eqnarray}
&& {P} (\hbox{mod}(L_1-x_1+1, L_1)+1,\cdots , \hbox{mod}(L_\nu-x_\nu, L_\nu)+1, \cdots ,
\nonumber \\
&&\quad\, \hbox{mod}(L_\rho-x_\rho, L_\rho)+1, \cdots , \hbox{mod}(L_d-x_d+1, L_d)+1 ;
\nu ,\rho)
\end{eqnarray}
in the original configuration ${\cal C}$. This is clearly a one--to--one
mapping from plaquettes to plaquettes in ${\cal C}$.
This completes the proof.

Then a configuration ${\cal C}$ with weight $\det D\exp(-S_{\rm g})$ 
transforms under {\sf CPT} into another configuration 
${\cal C}^{{\sf CPT}}$ with weight
$\left(\det D\right)^* \exp(-S_{\rm g})$ and the same Haar measure. 
This means that both configurations have the same odds to be selected by
an updating algorithm. If moreover we restrict our interest only to
{\sf CPT} invariant observables $O$ then we can say that the numerical
value $O[{\cal C}]$ appears with a probability proportional to 
$\exp(-S_{\rm g}[{\cal C}])\left(\det D[{\cal C}] + \left(\det D[{\cal C}]\right)^*\right)$.
Barring a factor 2 and dispensing with the {\sf CPT} transformed configuration
one can just count ${\cal C}$ with the weight $\exp(-S_{\rm g})\hbox{Re}\det D$.
This is the main result of our paper.

When we say {\sf CPT} invariant operators we mean that
the operator $O$ is {\sf CPT} invariant configuration by configuration.
Most of the operators are {\sf CPT} invariant after averaging over 
configurations (i.e. the corresponding observable is {\sf CPT} invariant),
but our constraint is stronger than this.

Focusing our attention only on {\sf CPT} invariant operators still allows 
to study several interesting problems. Indeed the average plaquette, chiral
condensate, Polyakov loop, etc. can be calculated by evaluating expectation 
values of {\sf CPT} invariant operators. On the other hand 
space--time valued operators $O(x)$ are in general not permitted because
our method requires that the equality $O(x)=O(-x)^\dagger$ holds 
configuration by configuration
which in general is not true. In particular all correlation functions are not
admissible.

Are there other discrete transformations such that the imaginary part of the
determinant is eliminated within groups of two configurations? 
We have performed the following test. We discretized
a 2--dimensional gauge theory\footnote{In 2 dimensions the action that introduces
the chemical potential on the lattice through the term $\mu\psi^\dagger\psi$ does
not present the problems discussed in~\cite{hasenfratzk}. The density of energy
obtained with this action on the lattice is $\mu^2/2\pi$, the same result
than in the continuum. However this is irrelevant for the purpose of 
the present test and we used the action in Eq.(\ref{action}).} on a
$2^2$ lattice and loaded it with an arbitrary configuration (an arbitrary $SU(3)$ matrix
on each of the 8 links). We calculated $\det D$ obtaining a complex number.
Then we rearranged the 8 $SU(3)$ matrices in all possible ways (8!=40320) and
for each permutation we 
placed them again on the 8 links and recalculated
the determinant. Only the transformation described in~(\ref{Ld}) yielded the
result that satisfies conditions~(\ref{conditions})
(permutations of the
8 links that can be viewed as spatial or temporal translations led to the same
result but they are clearly uninteresting). We conclude that there are no other
simple discrete transformations providing the complex conjugate of the
original configuration weight.

\section{A note on Monte Carlo simulation}

{} For the importance sampling to work it is necessary that the weight
be positive (it must behave as a probability). However Re $\det D$ can take
negative values and this fact limits the applicability of the method. By continuity
we expect that Re $\det D$ is positive in the majority of configurations when
$\mu$ is small enough. This suspicion is confirmed by numerical simulation.
In~\cite{toussaint} results from a simulation with 4 flavours 
of staggered fermions are presented.
A $4^4$ lattice is used at $\beta=4.8$ and $am=0.025$. In Fig.~1 we show the fraction
of configurations with a positive (negative) weight $N_+/N$ ($N_-/N$) 
as a function of $a\mu$ ($N\equiv N_++N_-$ is the total number of configurations).
It indicates that our
method can be used at moderate values of $\mu$. This may include the region in the
phase diagram $T$--$\mu$ where present and future heavy ion experiments (RHIC, LHC)
are going to be run (large $T$ and small $\mu$).

\begin{figure}[tbh]
\begin{center}
\includegraphics[width=0.8\textwidth]{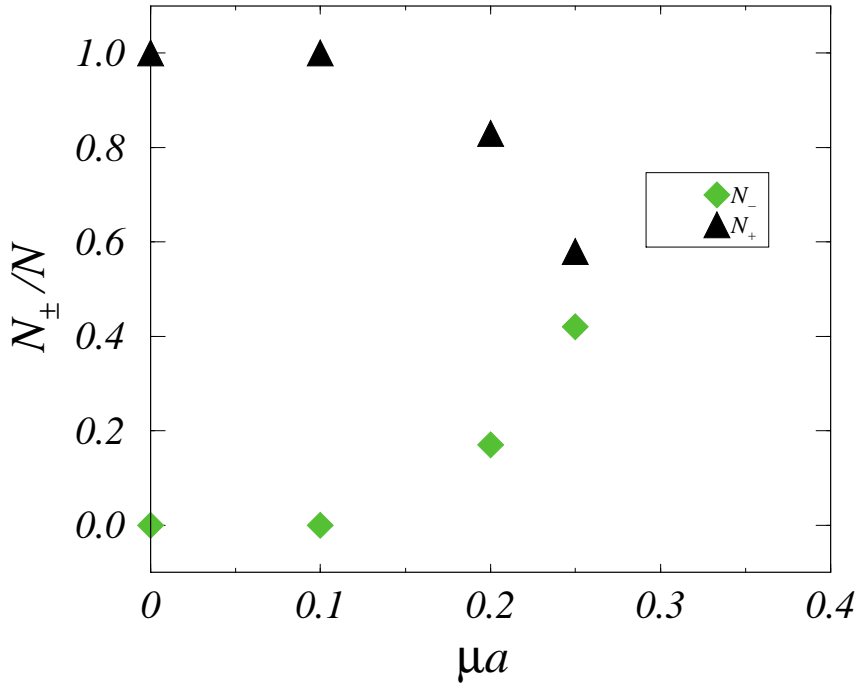}
\end{center}
\caption{\small{Fraction of positive and negative weighted configurations
as a function of the chemical potential. Data taken from~\cite{toussaint}.}}
\label{fig:4histo}
\end{figure}

Monte Carlo simulations with the weight Re $\det D$ would be greatly facilitated
if we were able to find a new matrix $\Delta$ such that Re $\det D=\det \Delta$ 
because then fast and well--known simulation methods for fermions~\cite{gottlieb}
could be used. We have not found a general and efficient algorithm to construct
the matrix $\Delta$ starting from $D$. Consequently we have to resort to algorithms
which explicitely calculate the determinant of $D$.

We shall not insist in these aspects of the problem as they will be analysed
in a future publication containing several numerical studies~\cite{prep}.

\section{Conclusions}

We have proved that the correct Boltzmann weight for updating full QCD in lattice
simulations at finite chemical potential is 
\begin{equation}
\exp\left(-S_{\rm g}\right)\hbox{Re }\det D
\label{we}
\end{equation}
where $S_{\rm g}$ 
is the pure gluon action and $D$ is the fermion matrix. We have shown that
this is true when we calculate expectation values of operators
that are {\sf CPT} invariant. For other operators our proof does not
work. Possibly in this case the assumption that the imaginary part
of the Boltzmann weight is eliminated by combining couples of configurations
should be relaxed.
Nonetheless the class of {\sf CPT} invariant operators include many
observables usually studied in the context of finite density systems.

Our algorithm has two problems which deserves further improvement: on one hand 
the weight~(\ref{we}) is not positive in general. We have shown that
it is mostly positive for moderate values of the chemical potential. On the other
hand the present method requires the explicit calculation of the determinant
of the fermion matrix $D$ which is very time consuming. In a future publication~\cite{prep}
we will give numerical results obtained by using Eq.(\ref{we}).

\vskip 1cm

\section{Acknowledgements}

It is a pleasure to thank G. Marchesini for useful discussions.

\vskip 5mm


\end{document}